\documentstyle[emulateapj]{article}
\def\gtorder{\mathrel{\raise.3ex\hbox{$>$}\mkern-14mu
             \lower0.6ex\hbox{$\sim$}}}
\def\ltorder{\mathrel{\raise.3ex\hbox{$<$}\mkern-14mu
             \lower0.6ex\hbox{$\sim$}}}
\def\ltsima{$\; \buildrel < \over \sim \;$}
\def\simlt{\lower.5ex\hbox{\ltsima}}
\def\gtsima{$\; \buildrel > \over \sim \;$}
\def\simgt{\lower.5ex\hbox{\gtsima}}
\def\rxte{{\it RXTE}}

\begin{document}

\submitted{To appear in The Astronomical Journal, October 2002} 
\title{X-ray vs. Optical Variations
       in the Seyfert~1 Nucleus NGC~3516:\\
 A Puzzling Disconnectedness}

   \author{Dan Maoz\altaffilmark{1,2}, Alex Markowitz\altaffilmark{3}, 
Rick Edelson\altaffilmark{3,4,5},
 \& Kirpal Nandra\altaffilmark{6,7}}

 \altaffiltext{1}{School of Physics \& Astronomy and Wise Observatory,
    Tel-Aviv University, Tel-Aviv 69978, Israel. dani@wise.tau.ac.il}

 \altaffiltext{2}{Astronomy Department, Columbia University, 
550 W. 120th St., New York, NY 10027}

 \altaffiltext{3}{Astronomy Department, University of California, 
Los Angeles, CA 90095-1562}

  \altaffiltext{4}{X-ray Astronomy Group, Leicester University, Leicester LE1 7RH, 
United Kingdom}

   \altaffiltext{5}{Eureka Scientific, 2552 Delmar Ave., Oakland, CA 94602-3017}

    \altaffiltext{6}{NASA/Goddard Space Flight Center; Laboratory for High
Energy Astrophysics; Code 662; Greenbelt, MD 20771}

    \altaffiltext{7}{Universities Space Research Association}

\begin{abstract}

We present optical broadband ($B$ and $R$) observations of the Seyfert
1 nucleus NGC~3516, obtained at Wise Observatory from March 1997 to
March 2002, contemporaneously with  X-ray 2--10 keV
measurements with \rxte. 
With these data we increase the
temporal baseline of this dataset to 5 years, more than triple to the 
coverage we have
previously presented for this object.
Analysis of the new data does not confirm the
100-day lag of X-ray behind optical variations, tentatively reported
in our previous work. Indeed, excluding the first year's data, which
drive the previous result, there is no significant correlation at any
lag between the X-ray and optical bands. 
We also find no correlation at any lag between optical flux and various
X-ray hardness ratios. We conclude that the close relation observed 
between the bands during the first year of our program was either a fluke,
or perhaps the result of the exceptionally bright state of NGC 3516 in 1997, to
which it has yet to return. Reviewing the results of published joint 
X-ray and UV/optical Seyfert monitoring programs, we speculate  that there
are at least two components or mechanisms contributing
to the X-ray continuum emission up to 10 keV: 
a soft component
that is correlated with UV/optical variations on timescales $\gtorder 1$ day, 
and whose presence can be
detected when the source is observed at low enough energies ($\sim 1 $keV),
is unabsorbed, or is in a sufficiently bright phase; 
and a hard component whose
variations are uncorrelated with the UV/optical.
 
\end{abstract}


\section{Introduction}

The paradigm that active galactic nuclei (AGNs) 
are powered by accretion onto massive black holes (MBHs)
has recently gained strong observational support,
with the
detection, in several AGNs, of X-ray emission lines that are thought to be 
broadened by relativistic effects near the MBH horizon
(Nandra et al. 1997; Sako et al. 2002), the
evidence for dormant black holes in many normal nearby galaxies
(Gebhardt et al. 2000; Ferrarese \& Merritt 2000), and 
the estimates of MBH masses in several tens
of AGNs via reverberation mapping (Kaspi et al. 2000).
 However, the detailed mechanisms
by which accretion produces the observed spectral energy distributions, 
as well as other properties, of AGNs are unknown, and observations
have placed few constraints on the many theoretical scenarios proposed.

It has been hoped that
flux variations in different energy bands would 
provide clues toward understanding the AGN emission processes.
In particular, a number of bright Seyfert-1 galaxies have been subject
to contemporaneous X-ray and UV/optical monitoring aimed at detecting
 inter-band lags, which could etablish a relation between emission
components, e.g., by identifying
the primary and secondary (i.e., reprocessed)
emissions (Done et al. 1990;  Clavel et
al. 1992; Kaspi et al. 1996; Crenshaw et al. 1996;
 Warwick et al.  1996; Edelson
et al. 1996; Nandra et al. 1998; Edelson et al.  2000;
Peterson et al. 2000;
Pounds et al. 2001; Turner et al. 2001; Collier et al. 2001;
Shemmer et al. 2001). 
However, the results of these programs, which have
searched for correlations and lags on timescales of hours to weeks,
have not been conclusive. It is generally true that UV/optical variation
amplitudes are much smaller than those in the X-rays, which could argue that
the X-rays are the primary emission. Clear 
lags between X-ray and UV/optical
variations have not been seen. In those
cases where correlation at a lag between different X-ray bands
 has been detected 
(sometimes with debatable
significance), the lag increased with band energy, indicating the X-rays
are secondary (e.g. Chiang et al. 2000). 
In a variant on the idea of searching for correlations
between fluxes at different bands, Nandra et al. (2000) found that the 
X-ray spectral index in NGC~7469 was correlated with UV flux at zero lag 
during a month-long campign on this Seyfert 1 galaxy. Papadakis, Nandra,
\& Kazanas (2001) have analyzed the cross-spectrum of variations in several
X-ray bands in this object, and found  that harder X-rays are delayed
with respect to soft ones, with the delay proportional to the Fourier period
probed. Such behavior is common in Galactic black hole binaries, but several
competing theoretical explanations exist for it.

The studies mentioned above have tended to be of limited duration - often just
a few days (Peterson et al. 2000 being the main exception).
A potential pitfall of short duration studies is that they may detect
few or no large-amplitude variation events with which to search for 
inter-band correlations. Furthermore, the results of variability studies
may depend on the timescale sampled, and different behavior may pertain 
to different sources. 
In 1997 we initiated a long-term X-ray/optical program to monitor continuously
several Seyfert 1 galaxies, 
such that month- and year-long variation timescales can
 be properly probed, as well as shorter timescales. X-ray observations
are obtained with the {\it Rossi X-ray Timing Explorer} (\rxte), and
optical data are from the Wise Observatory 1m telescope.

In Maoz, Edelson, \& Nandra (2000, hereafter Paper I), we presented
the first 1.5 year of X-ray and optical data for  
NGC~3516. Paper I found that the low-frequency component of the
X-ray variations appeared to mimic the optical variation during the
first year, but with a lag of $\sim 100$ days. However, this correlation
ceased in the last 6 months of the data. 
Paper I found that the
correlation was significant at the $\sim$99\% level, based on Monte Carlo
simulations that assumed power-law fluctuation power spectra with slope
--1.0 in the X-rays and --1.75 in the optical.  This was reasonable based
on the best information available at the time (Edelson \& Nandra 1999).
More recent data suggest a steeper X-ray power spectrum slope,
 of --1.35  (Markowitz \& Edelson 2002).  As discussed in Paper I, steeper
slopes will yield lower significance levels, and revised simulations will be
reported in a future paper (Edelson, Uttley, \& Markowitz 2002).
Although we proposed some 
physical explanations for the correlation, we cautioned that it
was driven by a single variation ``event'' and could therefore be a 
statistical coincidence. 
Here, we revisit NGC~3516 after having accumulated 5 years of
contemporaneous X-ray and optical data.  

\section{Optical and X-ray Light Curves}

The optical data presented in this paper span the period
from 1997, March 5, to 2002, March 7, and were obtained with the
Wise Observatory 1m telescope in Mitzpe Ramon, Israel. On the
nights when the galaxy was observed, Johnson-Cousins
$B$- and $R$-band images were obtained once per night.
We  used a $1024\times 1024$-pixel
thinned Tektronix CCD at the Cassegrain focus, 
with a scale of $0.7''$~pixel$^{-1}$. Exposure times were 3 min in $R$
and 5 min in $B$.
During this 1828-day period, useful data were obtained for 209 epochs
in $R$ and for 184 epochs in $B$.
The reader is referred to Paper I for details of the optical data reduction,
aperture photometry, and derivation of light curves relative to comparison
stars (up to six) in each frame.

Figure 1 shows the optical light curves we have obtained for NGC~3516.
In Figure 2 we plot on the same scale for each optical band the constant,
to within errors, 
 light curve of
one of the comparison stars, calculated relative to the other five stars.
As in Paper I, the $R$ and $B$ light curves of NGC~3516 in Figure 1
 show very similar variability patterns,
with peak-to-peak amplitudes of 0.5 mag and 0.9 mag, respectively.
The exact amplitude of the variations depends
on the choice of photometric extraction aperture, 
which will include a particular
fraction of stellar light from the galaxy. 
The above numbers
are therefore lower limits on the {\it intrinsic} variability amplitude
of the nucleus in each band, which is difficult to estimate.
Galaxy contamination is larger in the $R$ band, and at least some of the
difference in amplitude between $B$ and $R$ variations is due to this. 
In the last epoch presented here, the optical flux was at its lowest in the
past 5 years, and falling.

The observations and
reduction leading to the new \rxte~ data we present
 are as described in Edelson \& Nandra (1999),
 but using the most up to date background and calibration
files.
The \rxte~ data span the period from 1997, March 15 to 2002, February 26.
Up until 2000, Feb 18, the sampling interval between points was generally
about 4.3 days, except for several periods of more intense monitoring.
Then, after a 140-day period when \rxte~ did not observe NGC~3516, the
monitoring resumed and continued with a sampling interval of
about 17 days, except for a continuous 110~ks scan on 2001 April 10-11.

Since the $R$ and $B$ light curves are very similar, we will
refer to them collectively as ``the optical light curve''. We 
will use mainly the $R$ light curve, which is slightly better sampled
than $B$, in the figures and discussion below.
Figure 3 (top panel) shows again the $R$ light curve of NGC 3516,
but with a relative linear (rather than magnitude) flux scale.
The bottom panel shows
the \rxte~ X-ray (2-10 keV) light curve.
As found in Paper~I, which covered the first one-third of the baseline shown
in Fig. 3, the X-ray light curve has much larger variation amplitude
than the the optical, particularly at short timescales. In the last 3 years,
or so, the nucleus is faint relative to the first year, in the X-rays
as well as in the optical, if one ignores the occasional 
large X-ray flicker on small timescales.

As mentioned in \S1, a central and surprising result in Paper I was
an apparent optical-to-X-ray cross-correlation signal at $\sim 100$~day lag,
driven by the slow component of an outburst that appears in Figure 3 
between days 
600 and 900 in the optical, and between days 700 and 1000 in X-rays.
We revisit this issue now with our much-expanded dataset.
To isolate the relative contributions to the light curves and the correlations
made by fast and slow variations, we have smoothed the light
curves with a 30-day boxcar running mean. The smoothing was done
by replacing each observed flux with the mean of all fluxes that
are within $\pm 30$ days of it. In addition to the
unaltered light curves, we examine these 
smoothed versions and the residual light curves (i.e., the original
light curves minus their respective smoothed versions).

Figure 4 shows the smoothed X-ray and optical light curves superimposed.
To facilitate comparison of the two, all fluxes are plotted on a relative
linear scale, but the amplitude of the smoothed X-ray variations have
been scaled down by a factor of 4.
 The top
panel shows the two light curves with no lag, and the bottom panel shows
them with the X-rays advanced by 100 days. Clearly, the match
of the first year at 100-day lag does not persist or repeat in the new
data. Furthermore, a correlation at zero lag, noted in Paper I, and driven
by the simultaneous dip in the light curves around day 1000, does not
hold up. 
To study this question more quantitatively, and see if there are some other
lags at which any of the light curves are correlated, we have calculated    
the cross-correlation functions among the various light curves.
The z-transformed discrete correlation function (ZDCF;
Alexander 1997), a modification of the discrete correlation function
(Edelson \& Krolik 1988) was used.

Figure 5, top left panel, shows the cross-correlation between $B$ and $R$.
The high peak is at zero lag, confirming a result from Paper~I.
The top right
 panel shows the ZDCF between the residual (i.e., after subtraction
of a smoothed version) $B$ and $R$ light curves. The null correlation
between these close bands,
expecially around zero lag, suggests that
the short-timescale variations in the optical light curves (or at least in one
of them) are dominated by measurement error. This is not surprising, given
the estimated errors, and the small amplitude of the actual 
fast optical variations, as previously quantified in this object in  
more sensitive and densely
sampled {\it HST} data (Edelson et al. 2000).

The bottom left panel of Fig.~5 cross correlates the $R$-band and  
 2-10 keV light curves. A peak is seen at a lag of about 100 days,
in the sense that X-rays lag the optical. The bottom-right panel shows
the ZDCF for the smoothed optical and X-ray light curves. The 
peak correlation at 100-day lag is strengthened, indicating it is driven
by the slow components of the light curves. The position and height of
the ZDCF peaks in the latter two plots are very similar to those found
in Paper~I, based on the first
one-third of the data. Moreover, various other large
maxima and minima that appeared in these cross-correlations in Paper~I
do {\it not} appear in the present, expanded, dataset, 
arguing that those peaks,
at least, were artifacts of the sampling.

However, it is easy to confirm formally the visual impression from Fig. 4,
that the correlation at 100 days is still driven only by the first year's
outburst. Figure 6, top-left panel, correlates the smoothed X-ray and 
optical light curves, but excluding the first year's data. The
peak at 100 days is gone, and there is no clear and significant signal
at any lag. The top right panel of Fig. 6 shows that there is also no clear
correlation between the fast components of the $B$ and X-ray light curves,
as represented by the residual light curves. Note, however, that the point at
zero lag is the highest. Possibly there is, buried inside
the optical light curves, a positive correlation with the X-rays, which
could be recovered if the optical measurements had milli-magnitude accuracies,
rather than a few percent. Reaching such accuracies is probably unrealistic
in ground-based observations of a point source (the AGN) on a bright
galaxy background, but could be achieved with a photometrically
stable space telescope. On the other hand, Edelson et al. (2000) carried out
such an experiment (albeit limited to 3 days duration) and did 
not find such a correlation.

Finally, we investigate whether the optical and X-ray variations can 
be related via some observable other than the total flux. Following
the lead of Nandra et al. (2000, see \S1), we have calculated
the ZDCFs of the optical light curve vs  the ``softness'' ratio of counts
in  different X-ray bands. 
Testing among the various 
X-ray count ratios that can be formed from the 2-4 keV, 4-7 keV, and 7-10 keV
bands, their smoothed versions, and the various
optical light curves, we find no case of a clear correlation. For example,
the bottom panels of Fig. 6 show the ZDCFs of $R$ vs the 4-7 keV/ 7-10 keV 
count ratio, for both smoothed and unsmoothed light curves. Interestingly,
although there is no single clear peak, there is a fairly high correlation
plateau between about zero and 400 days lag. The source of this can be
seen in Fig.~7, which compares the optical light curve to the smoothed
4-7 keV/ 7-10 keV ratio curve. Both time series are plotted on a 
relative scale. One sees that, although there is no one-to-one
correspondence among the light curves, the X-ray spectrum was softer until 
March 1999, when the optical flux was generally high, than after May 1999,
when the optical flux was generally low. It is the lack of detailed
correspondence in this trend that washes out the correlation to a broad
plateau in the ZDCF. Naturally, we cannot say whether this trend is real, as it
is based not even on a full ``event'' (e.g. a rise and fall) in the light
curves. We also note that there is no analogous effect at lower X-ray
energies, as seen in the relative 2-4 keV/4-7 keV ratio plotted in the
bottom panel of in Fig. 7. In both of these softness ratio plots, the
typical errors on the ratios are 5-10\%. The large fluctuations in the
ratios are therefore significant, but some of the ``spikiness'' in the
latter parts of these
curves results from the 17-day sampling intervals during the last 600 days, 
and hence the modest effect of the  30-day smoothing.

\section{Discussion}

Much current thinking about the emission processes in AGNs
centers around the notion that the X-rays arise from very close
(within a few 
Schwarzschild radii) of
a massive black hole. Support for this idea
has come from the rapid variability that is observed in
X-rays (implying small physical scales), as well as the
detection in X-rays of a broad Fe K-shell emission line
in many Seyfert 1s (e.g., Nandra et al. 1997). The emission line is
thought to be gravitationally and Doppler
broadened fluorescence of the inner parts of an accretion disk, after
the disk is illuminated by the X-rays. More recently, such relativistic
emission lines from the Ly$\alpha$ transitions of several hydrogen-like ions
may have been detected in XMM-{\it Newton} data for two Seyfert galaxies 
(Branduardi-Raymont et al. 2001; Sako et al. 2002), though this claim has
been contested using {\it Chandra} data (Lee et al. 2001). 
The continuum-emission mechanism is not
known, but most commonly it is assumed that the X-rays are optical/UV photons
which have been upscattered by a population of hot electrons
(e.g., Sunyaev \& Truemper 1979).
The acceleration mechanism and geometry of the X-ray source is not
known. Neither is the source of seed photons, and despite some
substantial problems it is still usually assumed that the optical/UV
arises directly from an accretion disk (Shields 1978; Malkan 1983).
It has also been hypothesized that X--rays illuminating the disk,
or other optically thick gas, might be responsible for some or all
of the optical/UV radiation, via reprocessing (Guilbert \& Rees 1988;
Clavel et al. 1992).

Variability data
such as those we have presented above can provide constraints
on possible models. In summary of the observational results,
we
have found a similarity between the optical and X-ray
light curves during the first year of our program, when optically
the source was particularly bright, and with the X-rays
lagging  the optical variations by about 100 days. This correlation
disappeared in the last 4 years of the data, during which we see
no clear correspondence at any lag between the optical and the X-rays.
Furthermore we do not find any clear trends when we examine X-ray softness
ratios, rather than fluxes. The only positive signal we find are
a rough trend for a softer
spectrum in the 4-10 keV range when the source is optically brighter.
(The relation between X-ray spectral slope and brightness in {\it X-rays}
will be examined in a separate paper on the X-ray properties of this
object.)

Phenomenologically, the reality of any of these trends is debatable,
and all of them may be chance coincidences. A more stringent test must await 
the results of continued monitoring, during which NGC~3516 may perhaps 
recover to the high optical brightness 
it attained between mid-1997 and mid-1998.
The 
lack of any straightforward correlation between X-ray and optical 
fluxes, in its simplest interpretation, argues that there is no
physical relation between the emission in the two bands, except perhaps that 
both ultimately derive their energy from the central black hole.
If there is a
connection between the emission mechanisms in these two wavelength regimes, 
at the very least it must be complex enough to wash out any evidence for
it in the variability data.

 Is NGC~3516 
peculiar among AGNs in its lack of
a clear correlation between X-ray and optical/UV fluxes?
To address this, we
critically review the results of previous campaigns on this and other Seyfert
galaxies.\\
{\bf NGC~4051} Done et al. (1990) monitored NGC~4051 for 2 days, and 
found no correspondence
between the large-alplitude 2-10 keV variations seen with {\it Ginga}  
and the constant (to $< 1\%$) optical flux. Peterson et al. (2000) 
monitored this galaxy for 3 years with \rxte~ at 2-10 keV and with
ground-based optical spectroscopy. Typical sampling intervals were 1-2 weeks
in both wavelength regimes. In the third year, the source went into an
 extremely low X-ray state. While confirming the lack of correlation
found by Done et al. (1990) on short timescales, Peterson et al. (2000)
found that the
light curves are correlated at near-zero lag 
after smoothing on 30-day timescales. \\
{\bf NGC~5548} Clavel et al. (1992) observed 
NGC~5548 simultaneously with {\it Ginga} at 2-10 keV and with {\it IUE}
at 1350 \AA\ over a period spanning 51 days. The source brightness was
lower than average both in UV and in X-rays. The authors claimed a significant 
zero-lag correlation, 
yet this was based on nine epochs, and basically one-half of
an ``event'' in the light curves. 
Chiang et al. (2000) observed NGC~5548 for 2.8 days simultaneously with   
{\it EUVE} (0.14-0.18 keV), {\it ASCA} (0.5-1 keV), and \rxte~ (2-20 keV),
with 44 {\it EUVE} epochs. They found a good correlation between the three
bands, but as in the previous experiment on this object by 
Clavel et al. (1992), the correlation is dominated by a single ``step'' in
the light curves. The connection of the extreme-UV with the UV range was
previously given by Marshall et al. (1997) who compared {\it EUVE} measurements
to {\it IUE} and {\it HST} UV observations, but the correlation they claimed
was based on 10 data points spaced over 10 days, and a low correlation
coefficient.\\
{\bf NGC~4151} Edelson et al. (1996) combined 14 epochs of {\it Rosat}
1-2 keV data and four epochs of {\it ASCA} (0.5-1 keV) data 
(Warwick et al.  1996), and compared them to {\it IUE} ultraviolet
(Crenshaw et al. 1996), and Wise Observatory optical (Kaspi et al. 1996)
measurements
of NGC~4151 which were comtemporaneous over 10 days. The source was near
its peak historical brightness.
In this case, the light curves at all bands showed zero-lag similarities
on $\sim 1$-day timescales. However, the X-ray light curves had an
overall rising trend during the 10-day period, whereas a constant or falling
trend was seen in the UV and optical light curves. Thus, the X-ray to
UV/optical correspondence was far from perfect, and in some sense opposed.\\
{\bf NGC~7469}
Nandra et al. (1998) observed NGC~7469 for over a period of 30 days
with 30 epochs (after averaging) and found
that the \rxte~ 2-10 keV and {\it IUE} UV fluxes were poorly correlated.
Nandra et al. (2000) then found in these data a better correlation of the UV flux with
the X-ray slope, rather than X-ray flux. The object was close to its average
brightness in X-rays and in UV.\\
{\bf Akn 564} This narrow-line Seyfert 1 was monitored approximately
daily for 50 days in the optical (Shemmer et al. 2001), in the UV
with {\it HST} (Collier et al. 2001), and in the X-rays with {\it ASCA}
at 0.7-1.3 keV (Turner et al. 2001) and with \rxte~ at 2-10 keV (Pounds
et al. 2001). Although variation amplitudes in the UV and optical were
only of order a few percent, 
a  correlation at $<1$~day lag between UV and X-raya was 
reported by Shemmer et al. (2001), as well as a possible
 correlation between X-ray and optical, if only a particular segment of the
optical light curve, surrounding a relatively large event, is used in the
analysis. \\
{\bf NGC~3516} Edelson et al. (2000) monitored NGC~3516 continuously
for 3 days with \rxte~ and {\it ASCA} at 2-10 keV, and with {\it HST} in the
optical. They found no significant correlation between X-ray and optical
variations. Those observations took place on days 917-920 (see Fig. 4) when 
the source optical brightness was average but the X-ray flux was relatively
high. In the present work on this AGN, we find no correlation between
optical and 2-10 keV variations on timescales of days to 5 years,
except possibly a 100-day delayed correlation during 1997-1998, when the source
was extremely bright. We do not see a correlation with X-ray slope analogous
to that found in NGC~7469 by Nandra et al. (2000).   

If we now attempt to synthesize the above results, the following picture 
emerges. There have been several cases of little or no correspondence
between X-ray and UV/optical variability. There have also been several cases
where a correlation has been claimed, but the result is not conclusive due
to poor sampling, insufficient variability, or low significance. Perhaps
the most 
convincing flux correlation has been seen 
by  Edelson et al. (1996) between
soft (1-2 keV) X-rays and UV/optical flux in NGC~4151, yet, 
as mentioned above, the longterm
trends in the two bands were opposed, and no rigorous simulations
have been done to quantify the significance of the correlation. 
The correlation between UV flux
and X-ray slope found in NGC~7469 by Nandra et al. (2000) also seems secure.
These latter two results could arise if (but do not necessarily
imply that) the UV/optical is better
correlated with soft ($< 2$ keV)
X-ray variations than with the hard X-rays.
 It is also important
to note that, in terms of timescales, there has been no evidence in
any Seyfert 1 of
a relation between X-ray and UV/optical variations at short ($< 1$~day)
timescales, and all claimed correlations have been on $\ge 1$ day timescales.
The fast variations therefore appear be associated mainly with
the harder X-rays. This is supported by the finding by 
 that the X-ray variation power density spectrum
flattens with increasing energy  in NGC 7469 
(Nandra \& Papadakis 2001),
Akn 564, and Ton S180 (Edelson et al. 2001).

Why, then, is a
relation between optical flux and X-ray slope, such as seen in NGC~7469,
 not seen in NGC~3516 in
the present work? It can be argued that, contrary to NGC~7469,
NGC~3516 has strong and variable absorption in X-rays, and that this
variable absorption decorrelates an intrinsic relation in NGC~3516
that is similar to the one in NGC~7469. Evidence for this can be seen in 
the fact that, in the present data for NGC~3516, the 4-7 keV and 7-10 keV
light curves, as well as the full 2-10 keV light curve are more similar 
to each other than to the 2-4 keV light curve, where absorption will
be strongest.

On the other hand, the X-ray 
absorption  in NGC~3516 is comparable to that in NGC~4151, where a UV-X-ray 
flux correlation is seen in a band that is only slightly softer than the
\rxte~ band. Perhaps this objection can be overcome by noting that the 
NGC~4151 correlation was seen when this source was exceptionally bright. 
If the source brightness is a factor, it can further
be argued that a flux correlation was indeed seen in NGC~3516, but only
during the first year of our program, when the source was exceptionally
bright, as was the case in NGC~4151. Indeed, we note in NGC~3516 
that the 2-4/4-7 keV and  4-7/7-10 keV ratios
and the total X-ray counts all seem to track each other better during the
first 700 days, when the source was brighter. Source brightness could
conceivably affect the correlations by
 making visible the high-energy 
tail of the actually correlated emission at low energies, or by
ionizing the absorbing gas, and thus reducing the decorrelating effect
of the variable absorption. The clear change in X-ray spectral softness
around day 1300 and the  accompanying optical dimming, while not
necessarily connected, are at least  consistent with the expectation
that a more photon-starved corona will produce a harder spectrum.
Alternatively, rather than source brightness playing a role, it may be
intrinsic differences between objects. For example, the fast, uncorrelated
X-ray emission may be always dominant in NGC~3516, and hence swamp
out the soft correlated emission in the light curves.

According to this picture, then, at least two components contribute to
the X-ray continuum emission of Seyfert nuclei: a soft component which 
is temporally related to the UV/optical continuum, and which can be
discerned in $> 1$~keV variability data only when the source is bright enough
or relatively unabsorbed; and a fast/hard component that 
varies independently. We note that by ``components'' we do not necessarily mean
emission from physically distinct regions (e.g., Shih, Iwasawa, \& Fabian
 2002).  Instead, the two components that contribute to the X-rays could
arise from the same region via different mechanisms. For example, 
the physical
conditions in the coronal regions could be affected autonomously by
two processes, such as
by time evolution (e.g. Poutanen and Fabian 1999) and 
by changes in seed photon input which cause an overall temperature change in
all the coronal regions. One or the other
of these mechanisms could dominate in a particular source at a particular time
and on a particular timescale.

  This empirical two-component picture and its underlying drivers
could be tested via a monitoring
program using X-ray observations with sufficient spectral
resolution and signal-to-noise ratio to disentangle the continuum from the
absorption, and measure the variability of the intrinsic flux and spectrum.
Another path is to search for the the spectral-slope/optical-flux
correlation in other objects, having either strong or weak X-ray absorption. 
We intend to do this in future papers for NGC~4151, NGC~5548, and PG0804+762.

Future programs can also test an alternative interpretation of all previous
results, namely, that there is no real correlation of any sort between
variations in optical/UV and X-ray bands. In that case, even the more
convincing correlations seen, such as Edelson et al. (1996) and Nandra et
al. (2000) are chance coincidences arising in the comparison of 
unrelated red-noise light curves. 
The simulations testing for this possibility can also be refined.
For example, the simulations in Paper I
assumed an X-ray power spectrum slope of $-1$, 
but  recent work indicates a slope of
$-1.35$ or steeper over the time scales of interest (Markowitz \& Edelson
 2002),
which would lead to a lowered significance level.
The simulated light
curves can include amplitude randomization (Timmer \& Koenig 1995) as 
well as phase randomization.
We do emphasize that all future claims of inter-band correlations
would benefit by simulations demonstrating their significance.

\acknowledgements

We would like the thank the following observers at Wise,
who contributed their efforts and observing time to obtain
the data presented here:  R. Be'eri, T. Contini,
J. Dann, A. Gal-Yam, U. Giveon, A. Heller, S. Kaspi, Y. Lipkin,  I. Maor,
H. Mendelson, E. Ofek,  A. Retter, O. Shemmer,
 G. Raviv, and S. Steindling. We are also grateful for the assistance
of the Wise Observatory staff: S. Ben-Guigui, P. Ibbetson, and E. Mashal.
 T. Alexander is
thanked for providing his ZDCF code, and the anonymous referee for 
useful comments.

\begin{figure}
\epsscale{0.75}
\plotone{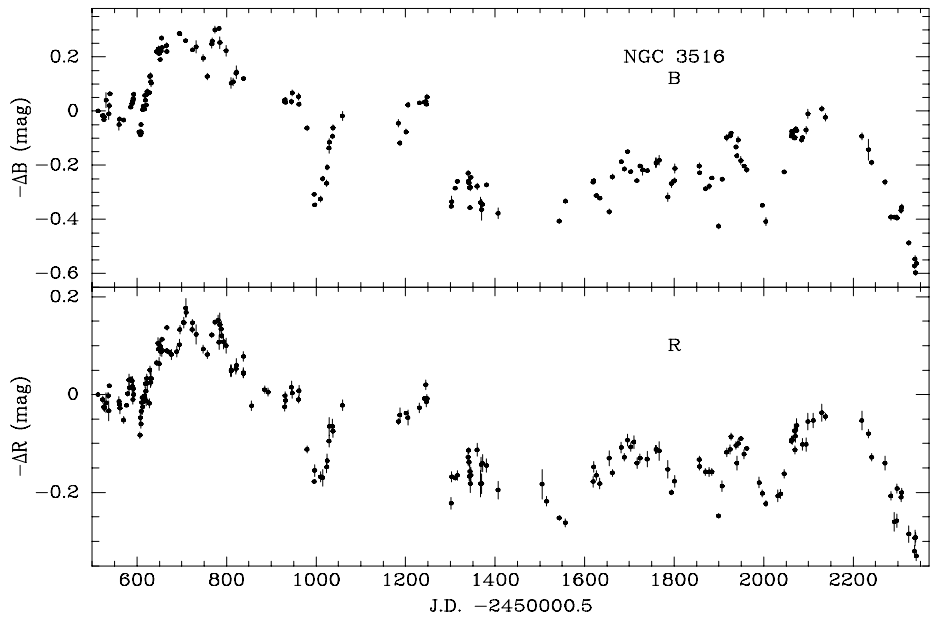}
\caption{$B$-band (top panel) and $R$-band (bottom panel)
 light curves for NGC~3516. 
}
\end{figure}

\begin{figure}
\epsscale{0.75}
\plotone{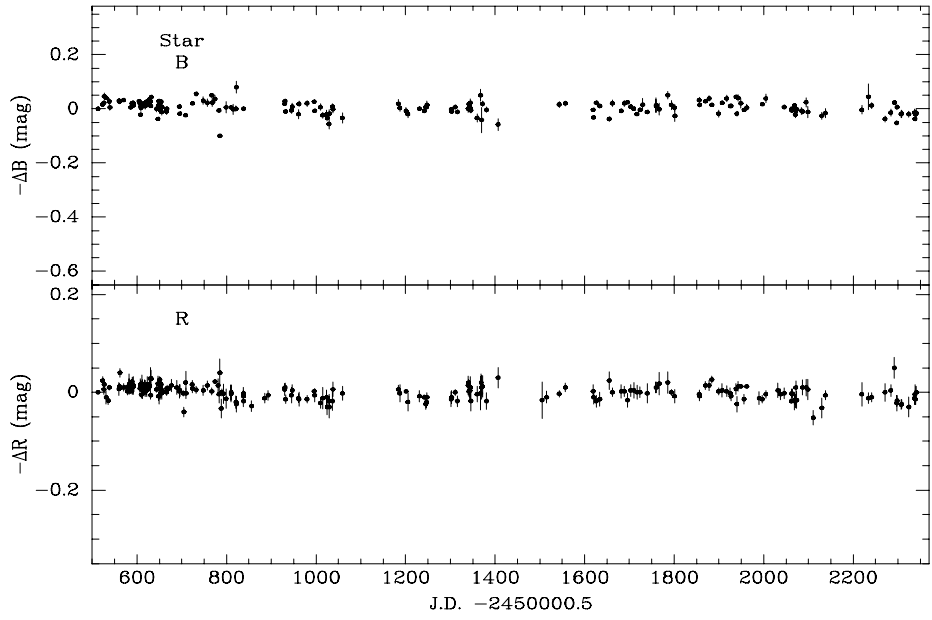}
\caption{$B$-band (top panel) and $R$-band (bottom panel)
 light curves for one of the comparison stars, measured in the same way
as the Seyfert nucleus. 
}
\end{figure}

\begin{figure}
\epsscale{0.75}
\plotone{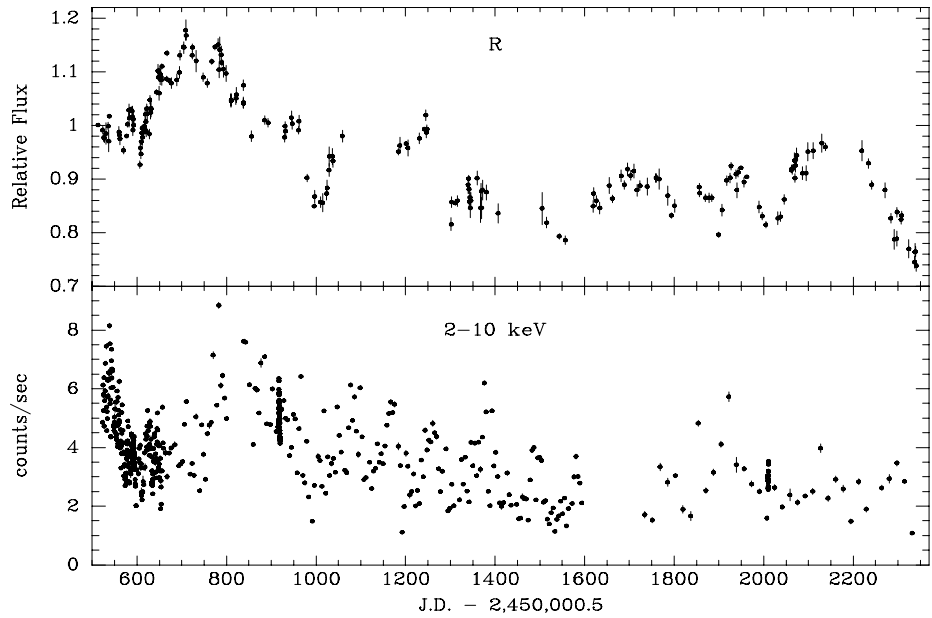}
\caption{Top panel: $R$-band light curve of NGC 3516,
but with linear relative flux scale.
Bottom panel:
 \rxte~ X-ray (2-10 keV) light curve.
}
\end{figure}

\begin{figure}
\epsscale{0.75}
\plotone{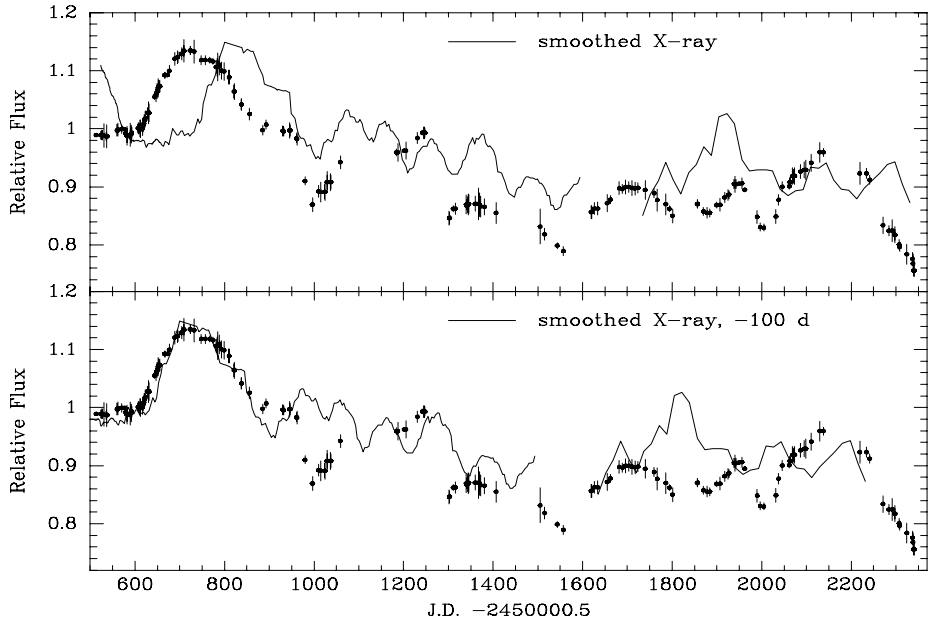}
\caption{$R$-band light curve (boxes) and
X-ray light curve (solid line), both after smoothing
with a 30-day boxcar running mean.
All fluxes are plotted on a relative
linear scale, but the amplitude of the smoothed X-ray variations 
is scaled down by a factor of 4. 
Top
panel shows the two light curves with no lag, and  bottom panel shows
them with the X-rays advanced by 100 days. Note how the excellent match
of the first year at 100-day lag does not persist.}
\end{figure}

\begin{figure}
\epsscale{0.75}
\plotone{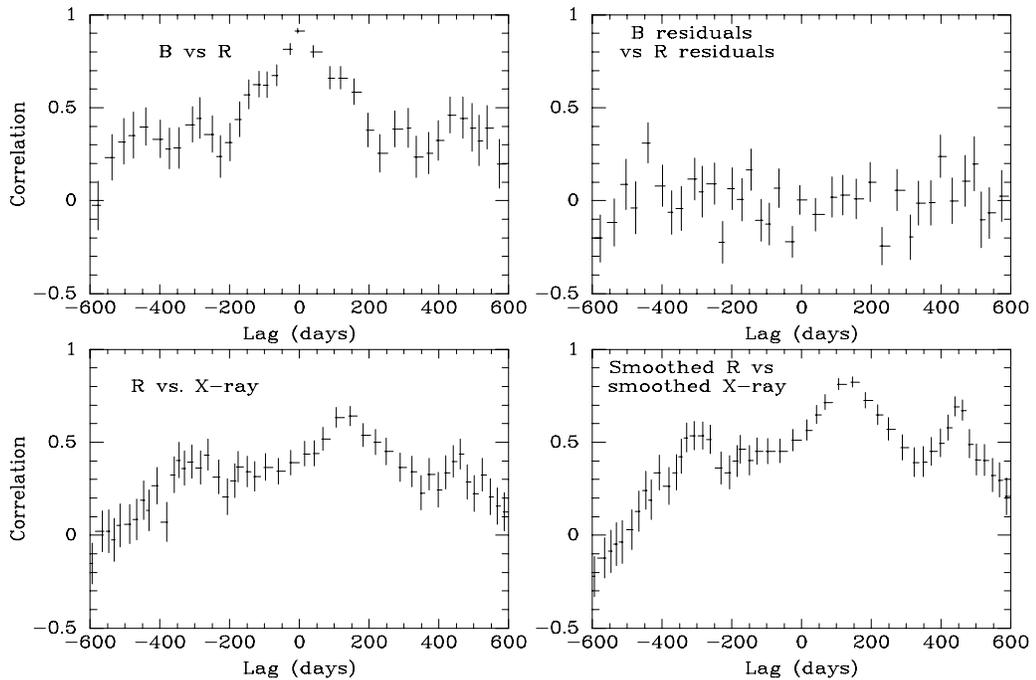}
\caption{ZDCF cross-correlations between various light curves,
as labeled. ``Smoothed'' refers to light curves that have been 
smoothed with a 30-day boxcar running mean, and ``residuals''
refers to the difference between an original light curve and
its smoothed version.
}
\end{figure}

\begin{figure}
\epsscale{0.75}
\plotone{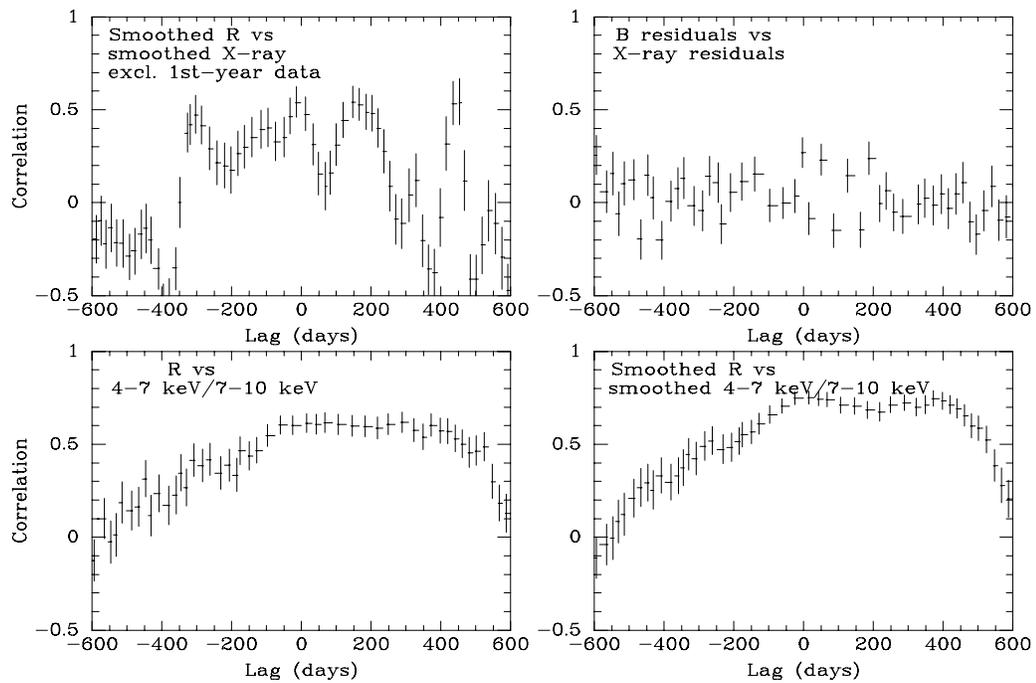}
\caption{Additional cross-correlations, as in Fig. 5.
Top left panel shows how the exclusion of the first year's
data removes the high correlation at any clear lag between
the X-rays and the optical. In the bottom panels, the $R$ light
curve is cross-correlated with the ``softness ratio'' of counts
in the 4-7 kev and 7-10 keV bands, and with its temporally
smoothed version.
}
\end{figure}

\begin{figure}
\epsscale{0.75}
\plotone{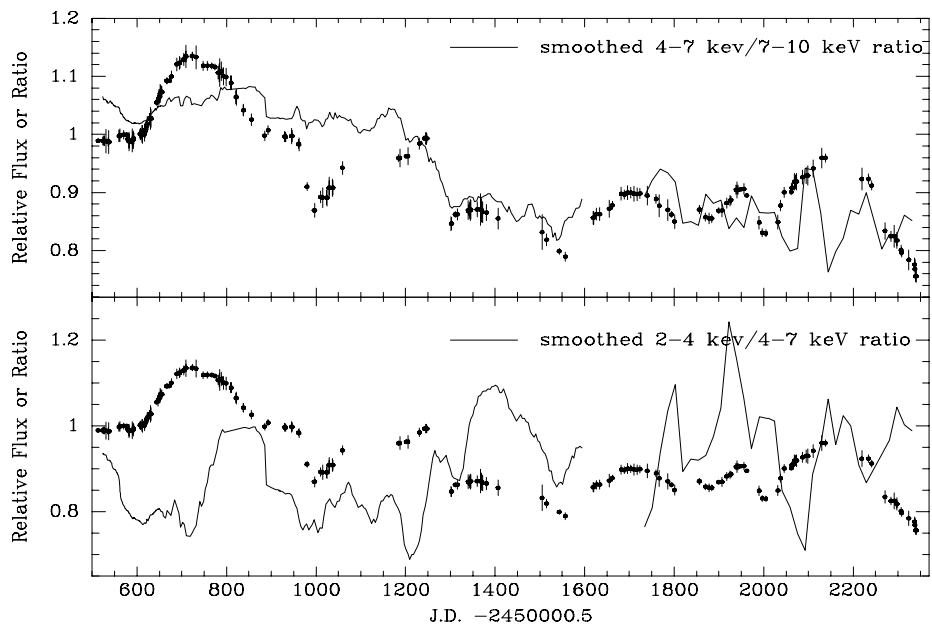}
\caption{Top panel:
Comparison of the smoothed $R$ light curve (boxes) to the smoothed
4-7 keV/ 7-10 keV count ratio vs. time (solid line). To ease comparison,
both time series are plotted on a relative 
scale normalized to unity. 
Bottom panel: same, for  the  2-4 keV/4-7 keV count ratio. 
}
\end{figure}

\end{document}